\documentclass[10pt]{article}
\usepackage{amsmath,amssymb,amsthm}
\usepackage{graphicx}
\usepackage{cite}
\title{Lateral Casimir Force between Two Sinusoidally Corrugated Eccentric Cylinders Using Proximity Force Approximation}
\author{M. R. Setare, A. Seyedzahedi \\
\small{\emph{Department of Science, University of Kurdistan, Sanandaj, Iran.}}\\
\small{Email address: \emph{rezakord@ipm.ir}} }

\date{}

\theoremstyle{plain}

\begin{document}

\maketitle

\setlength{\parindent}{0pt}

\begin{abstract}
\noindent
This paper is devoted to the presentation of the lateral Casimir force between two sinusoidally corrugated eccentric cylinders. Despite that applying scattering matrix method explains the problem exactly, procedure of applying this method is somehow complicated specially at nonzero temperature. Using the proximity force approximation (PFA) helps to achieve the lateral Casimir force in a truly explicit manner. We assume the cylinders to be slightly eccentric with similar radiuses and separations much smaller than corrugations' wave length for the validity of PFA. For such short distances the effect of finite conductivity would be non negligible. In addition to the effect of finite conductivity, we investigate thermal corrections of the lateral Casimir force to reduce the inaccuracy of the result obtained by PFA. Assuming the Casimir force density between two parallel plates, the normal Casimir force between two cylinders is obtained. With the aid of additive summation of the Casimir energy between cylinders without corrugation, we obtain the lateral Casimir force between corrugated cylinders.

\emph{keyword:}  Lateral force, Corrugated cylinders, Finite temperature.

\end{abstract}

\section{Introduction}
The Casimir force \cite{casimir} between two neutral conductors arising from the modification of the zero point energy associated with the quantum fluctuations, has got very fast theoretical and experimental advancement in the last decade \cite{Milonni,Milton,Advances,Barrera}. The pioneer studies of the Casimir effect have been done for two parallel plates or plate-sphere geometries. In $2006$ Dalvit and his collaborators have analyzed a different geometry. They have considered two eccentric real metal cylinders \cite{Dalvit1} and by using the proximity force approximation they have obtained the Casimir interaction energy. Nevertheless, afterwards they have investigated an exact solution for the mentioned geometry \cite{Dalvit2}. Based on the scattering matrix methods Casimir calculations for sphere-plate geometries has been extremely done since 2007 \cite{Emig3,Emig4,Neto1,Canaguier}. In these investigations the plate is considered to be perfectly conducting whereas on the sphere the electromagnetic field satisfied either perfect conductor boundary conditions or a dielectric with constant. The Casimir-Polder forces between an atom and a surface with arbitrary uniaxial corrugations has been presented applying a technique that is fully nonperturbative in the height profile of the corrugation \cite{0810.3480}. And by using height distribution function, effect of roughness or surface modulations on the distance dependence of power-law interactions between curved objects at proximity has been investigated recently in \cite{1303.2499}. Applying worldline numerics \cite{Langfeld1,Langfeld2}, the Casimir interaction energies for the sphere-plate and cylinder-plate configuration due to the scalar-field fluctuations under the Dirichlet boundary conditions for a wide range of curvature parameters has been examined in \cite{0601094}. Based on a high-precision calculation applying worldline numerics, Gies and Klingm$\ddot{u}$ller have concluded the validity bounds of the proximity force approximation quantitatively. They observed that for the accuracy goal of $1\verb"%"$, the PFA is valid for $s/r\leqslant 0.00755$. Where $s/r$ is called the curvature parameter, $r$ is radius of the sphere at a minimal distance $s$ from a plate. Analysis shows that, using PFA to obtain plate-sphere result from the plate-plate conclusion for corrugations with wave length $\lambda$ is a good approximation until $r s \gg \lambda^2$ and the amplitude of corrugation is smaller than the other length scales (see \cite{0603120} and references there in).

 Calculations of the Casimir force between a plane and a nanostructured surface at finite temperature in the framework of the scattering theory has been presented in \cite{1212.5479}. Considering regularized zero point energy density for two parallel plates, a derivation for the normal Casimir force between two parallel eccentric cylinders has been presented  \cite{1210.6813}. The normal Casimir force as the well known Casimir force acts perpendicular to the surfaces. When the interacting bodies are located asymmetrically or they are not isotropic, a lateral Casimir force may exist. Especially, two sinusoidally corrugated surfaces experience a lateral Casimir force \cite{Chiu,Chen,Emig2,Golestanian}. Theoretical predictions of the lateral Casimir force and its corresponding torque were performed in \cite{Parsegian, Barash}. This lateral force acts tangentially to the surface. It has been investigated that the miniaturized parts of nanoscale devices may couple by this lateral force without any contact\cite{Emig3,Miri2,Nasiri}. Intuitively one expects this lateral Casimir force to be a solution to the problems like friction, adhesion and wear in nanomachines. Sinusoidally corrugated plate and cylinder, considered as rock and pinion, has been introduced as a mechanical rectifier in \cite{Ashourvan,Moradian}. Rock and pinion are suppose to be coupled by the quantum fluctuations. In a recent investigation angle dependence of the Casimir force between corrugations has been demonstrated \cite{Zandi}. The lateral Casimir force between a sinusoidally corrugated sphere and plate has been studied experimentally and complete measurement data has been presented in \cite{Chiu}. The mentioned experimental and theoretical studies illustrate the realization of such kinds of nanoscale devices.
\par


In an attempt along these lines, we pay attention to the lateral Casimir force between two sinusoidally corrugated eccentric cylinders. Considering recent progress in nanotechnology, these sinusoidally corrugated cylinders may play an important role just like gears without contact in the micro-electromechanical systems (MEMS). Leading and next to leading order Casimir torque between concentric corrugated cylinders were derived for the scalar case in Dirichlet and Weak coupling limit \cite{Pelaez1} and similar investigation has been done for corrugated surfaces (see \cite{Pelaez2} for example). This kind of investigations lead in an exact result, but there are some difficulties in applying the scattering matrix method to obtain exact result. A comparison has been done between the exact result and the one obtained from the PFA in the weak coupling limit \cite{Pelaez3}, it has been noted that PFA is a good approximation for separations much smaller than corrugations' wave length. It is well known that for corrugated surfaces the use of the proximity force approximation may not lead to precise expressions for the Casimir force, but using this approximation leads in a simple procedure to achieve our purpose and removes difficulties appear in the exact investigations. We assume the cylinders to be slightly eccentric with similar radiuses for the validity of PFA. For the short distances the effect of finite conductivity would be non negligible. We also consider the finite-temperature corrections of the Casimir force to reduce the inaccuracy of the result obtained by PFA. In order to obtain the lateral Casimir force organization of this paper is as follows, in sec. 2. considering regularized zero point energy density for two parallel plates, we present a derivation for the normal Casimir force between two parallel eccentric cylinders. In sec. 3. we obtain the Casimir energy density for corrugated cylinders by the additive summation of the results obtained for cylinders without corrugation.
Considering finite conductivity of metals, we obtain the corresponding corrections in the Casimir interaction energy. Then we investigate the finite temperature corrections of the lateral Casimir force in sec. 4. Assuming thermal correction of the Casimir force density between two perfect conductor plates and with the aid of PFA we obtain the finite temperature lateral Casimir force in the high temperature limit.

\label{}

\section{The Casimir Interaction Energy between Two Eccentric Cylinders }
The scalar Casimir energy per unit area for two parallel perfect conductor plates with separation distance $H$ and Dirichlet-Dirichlet or Neumann-Neumann boundary condition is given by $E_{pp}(H)=- \frac{\pi^{2} \hbar c}{1440 H^{3}}$  \cite{casimir,Milonni,Bordag}. This leads in the normal Casimir force density $\mathfrak{F}_{pp}(H)=- \frac{\pi^{2} \hbar c}{480 H^{4}}$. The electromagnetic case is obtained from the scalar one by a factor $2$, for this simple geometry. Now consider two slightly eccentric cylinders with radiuses larger than separation distance between them. Assume that cylinders are corrugated and the following functions describe their longitudinal corrugations \cite{Chen}
\begin{eqnarray} \label{corrugation}
r_{1}=A_{1} \sin({2 \pi x}/{\lambda}) \qquad  \\\nonumber
r_{2}=h+A_{2} \sin({2 \pi x/}{\lambda}+\varphi)
\end{eqnarray}

\begin{figure}[h]
\hspace{+2 cm}\includegraphics [width=0.65 \columnwidth] {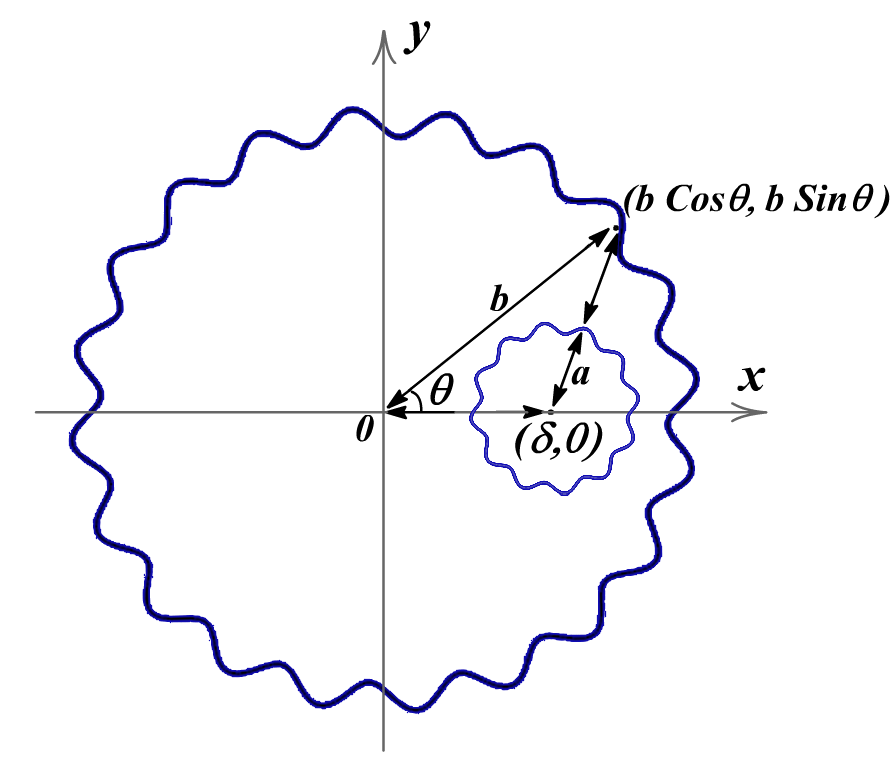}
\caption{The cross section of two parallel eccentric cylinders. Distances are overstated in the figure. Cylinders are slightly eccentric and $a\simeq b$.}
\label{cylinders}
\end{figure}

where $x$ denotes the coordinate, ${2 \pi x/}{\lambda}$ is an angle, $\varphi$ is the phase shift, $A_1$ and $A_2$ are the corrugation amplitudes, $h$ is the mean separation distance between the two surfaces of cylinders and $\lambda$ is the corrugation wavelength. Therefore the shortest separation distance between two points of the corrugated surfaces can be written as

\begin{eqnarray} \label{separation}
r_{2}-r_{1}= h + \beta\cos({2\pi x}/{\lambda-\alpha)})
\end{eqnarray}

where $\beta$ and $\alpha$ are both functions of $\varphi$ as follows:
\begin{eqnarray} \label{functions}
\hspace{-1 cm}&& \beta(\varphi)=(A_1^2+A_2^2-2 A_1 A_2 \cos\varphi)^{1/2} \\ \nonumber
\hspace{-1 cm}&&\alpha(\varphi)=(A_2 \cos\varphi-A_1)/(A_2 \sin\varphi).
\end{eqnarray}

As it has been investigating in \cite{Teo} the normal Casimir force between two parallel eccentric cylinders is obtained by integrating the Casimir force density over the area of one of the cylinders, i.e.
 \begin{eqnarray} \label{force}
\hspace{-1 cm}F_{cc}=\int  \mathfrak{F}_{pp}(H) ds,
\end{eqnarray}
considering that $H$ is the distance from a point of the integrated cylinder to the other one. They have used polar coordinates to determine that
the shortest separation distance from the point with parameter $\theta$ from the integrated cylinder (one of radius $b$) to the cylinder of radius $a$ is $\sqrt{b^2+\delta^2-2 b  \delta \cos\theta}-a$ where $\delta=b-a-h$.

\begin{eqnarray} \label{normal}
\hspace{-1 cm}F_{cc}=-\frac{\pi^2 b L}{240}\int_0 ^\pi \frac{1}{(\sqrt{b^2+\delta^2-2 b \delta \cos\theta}-a)^4} d\theta,
\end{eqnarray}

where $L$ is length of the cylinders and we focus on the special case of $a\simeq b$ in which the Casimir force between slightly eccentric cylinders may be obtained. Applying the change of variables (as hinted in \cite{Teo}) $u =\frac{\sqrt{b^2+\delta^2-2 b \delta \cos\theta}-a}{h}$, in the limit $h\rightarrow0$ the author has found the leading order term of the proximity force approximation of the configuration as follow
\begin{eqnarray} \label{leading}
\hspace{-1 cm}F_{cc}\thicksim-\frac{\pi^3 \sqrt{a b} L}{768 \sqrt{2(b-a)} h^{7/2}}.
\end{eqnarray}
One may obtain the Casimir energy between two interior eccentric cylinders without corrugations by integrating the normal Casimir force with respect to separation distance $h$,

\begin{eqnarray} \label{E-cc}
\hspace{-1 cm}E_{cc}=\int_{h}^\infty F_{cc} dh =\frac{2}{5} \bigg(\frac{\pi^3 \sqrt{a b} L}{768 \sqrt{2(b-a)}h^{5/2}}\bigg ).
\end{eqnarray}

\section{The Casimir Energy Density for Two Corrugated Eccentric Cylinders}

Considering the equal probability of all the separation distances $r_2-r_1$ introduced by Eq. (\ref{separation}) \cite{Chen, Bordag, Emig, Klimchitskaya}, with respect to the additive summation of the results obtained for cylinders without corrugation the Casimir energy density For corrugated cylinders with large corrugation wavelength $\lambda > r$ can be found

\begin{eqnarray} \label{E-corrugated}
& \hspace{-4 cm} E_{cc}^{cor} =\frac{1}{\lambda}\int_{0}^\lambda E_{cc}(r_2-r_1) dx \nonumber \\
&= \frac{2}{5 \lambda} (\frac{\pi^3 \sqrt{a b} L}{768 \sqrt{2(b-a)}})\int_{0}^\lambda \frac{1}{(h + \beta \cos({2\pi x}/{\lambda-\alpha)})^5/2}  dx.
\end{eqnarray}

The lateral Casimir force is obtained by taking derivative with respect to the phase shift
\begin{eqnarray} \label{F-lateral}
 F_{cc}^{lat} = - \frac{2 \pi}{\lambda} \frac{\partial}{\partial \varphi} E_{cc}^{cor} =
  \frac{- \pi ^3 \sqrt{a b} L}{960 \sqrt{2(b-a)}\lambda} \frac{A_1 A_2 \sin\varphi}{\sqrt{A_1^2+A_2^2-2 A_1 A_2 \cos\varphi}}\Pi(h,\alpha,\beta)
\end{eqnarray}
where $\Pi(h,\alpha,\beta)$ is
\begin{eqnarray} \label{pi}
\frac{1}{6 \beta (-\beta + h)^3 (\beta +h)^{5/2} \pi }
\bigg \{ h (29 \beta^2 + 3 h^2)\bigg( \mathcal{E}\big(\pi -\frac{\alpha}{2}, \frac{2 \beta}{\beta+h}\big)+ \mathcal{E}\big(\frac{\alpha}{2},\frac{2 \beta}{\beta+h}\big)\bigg)\\ \nonumber +
     (\beta - h) (5 \beta^2 +3 h^2)\bigg( \mathcal{F}\big(\pi -\frac{\alpha}{2},\frac{2 \beta}{\beta+h}\big) +\mathcal{F}\big(\frac{\alpha}{2},\frac{2 \beta}{\beta + h}\big) \bigg )\bigg \},
\end{eqnarray}
and $\mathcal{F}$ and $\mathcal{E}$ are the elliptic integrals of the first and second kind respectively.
\\
\par
Applying the proximity force approximation for the case of two exterior cylinders (at small separation), for the normal Casimir force is obtained \cite{Teo}

\begin{eqnarray} \label{exterior}
\hspace{-1 cm}F_{cc}\thicksim-\frac{\pi^3 \sqrt{a b} L}{768 \sqrt{2(b+a)} h^{7/2}}.
\end{eqnarray}
Following steps just like the case of two cylinders out of each other one can easily obtain a similar result for the lateral Casimir force corresponding to the case of two exterior cylinders
\begin{eqnarray} \label{similarly}
 F_{cc}^{lat} =
  \frac{ \pi ^4 \sqrt{a b} L}{960 \sqrt{2(b+a)}\lambda} \frac{A_1 A_2 \sin\varphi}{\sqrt{A_1^2+A_2^2-2 A_1 A_2 \cos\varphi}}\Pi(h,\alpha,\beta),
\end{eqnarray}
where $\Pi(h,\alpha,\beta)$ is introduced before in Eq. (\ref{pi}). It is worth mentioning that one can obtain this result just by replacing $(b-a)$ in Eq.(\ref{F-lateral}) by $(b+a)$.
\par
Considering finite conductivity of metals, corresponding corrections may be included in the mentioned scalar Casimir energy per unit area for two parallel perfect conductor plates \cite{Dzyaloshinskii,Schwinger,Bezerra}
\begin{eqnarray} \label{delta}
E_{pp}(H)=- \frac{\pi^{2}}{1440 H^{3}} \bigg(1+ \sum_{n=1}^{4} C_n \big(\frac{\lambda_p}{2 \pi H}\big)^n\bigg),
\end{eqnarray}
where $\lambda_p$ is the plasma wavelength and the coefficients $C_n$ are as follows
\begin{eqnarray} \label{C}
C_1 = -4, C_2 = \frac{72}{5}, C_3=-\frac{320}{7}(1-\frac{\pi^2}{210}), C_4=\frac{400}{3}(1-\frac{163 \pi^2}{7350}). \nonumber
\end{eqnarray}
At the separations $H \geq \lambda_p$, Eq. (\ref{delta}) is applicable. Imposing PFA approximation on the correction terms leads in the following correction in the force derived by Eq. (\ref{leading})
\begin{eqnarray} \label{conductivity}
\hspace{-1 cm} \Delta F_{cc}\thicksim \sum_{n=1}^{4}\frac{n}{h^ {9/2+n}}\,B_n \,(\frac{\lambda_p}{2 \pi})^{n},
\end{eqnarray}
where
\begin{eqnarray} \label{B_n}
B_n =-\frac{\pi^2 L \sqrt{ab}}{240 \sqrt{2(b-a)}} \frac{ \sqrt{\pi}\;\Gamma(9/2+n)}{\,\Gamma(5+n)} C_n,
\end{eqnarray}
in which $C_n$ are the coefficients introduced before. By integrating the correction of the Casimir force with respect to the separation distance $h$, effect of the finite conductivity in the Casimir energy is
\begin{eqnarray} \label{conductivity2}
\hspace{-1 cm} \Delta E_{cc}\thicksim - \sum_{n=1}^{4}\frac{n}{(7/2+n)h^ {7/2+n}}\,B_n \,(\frac{\lambda_p}{2 \pi})^{n}.
\end{eqnarray}
Considering the mentioned sinusoidally corrugations with large corrugation wavelength by imposing additive summation of Eq.(\ref{E-corrugated}) on the corrections, effect of the finite conductivity on the corrugated Casimir energy can be obtained
\begin{eqnarray} \label{conductivity3}
\hspace{-2.5 cm} \Delta E_{cc}^{cor}\thicksim -\frac{\pi^2 L \sqrt{ab}}{240 \sqrt{2(b-a)}} \big(\frac{\lambda _p}{2 \pi}\big)\frac{\sqrt{h+\beta}}{(h^2-\beta^2)^4}\bigg[ -\frac{1}{480}\Xi_1+\frac{3}{800}\big(\frac{\lambda _p}{2 \pi}\big)\frac{1}{(h^2-\beta^2)}\Xi_2 \\ \nonumber
+ \frac{1}{2352}(1-\frac{\pi ^2}{210})\big(\frac{\lambda _p}{2 \pi}\big)^2 \frac{1}{(h^2-\beta^2)^2}\Xi_3+
\frac{5}{16128}(1-\frac{163 \pi ^2}{7350})\big(\frac{\lambda _p}{2 \pi}\big)^3 \frac{1}{(h^2-\beta^2)^3}\Xi_4\bigg],
\end{eqnarray}
with the set of functions $\Xi_i$, $i=1,...,4$, functions of $ (h,\alpha ,\beta)$, as the follows:
\begin{eqnarray} \label{conductivity4}
\Xi_1=16h(11h^2+13\beta^2)\bigg( \mathcal{E}\big(\pi -\frac{\alpha}{2}, \frac{2 \beta}{\beta+h}\big)+ \mathcal{E}\big(\frac{\alpha}{2},\frac{2 \beta}{\beta+h}\big)\bigg)\qquad\qquad\qquad\qquad\qquad\qquad\qquad\quad \\ \nonumber
-(h-\beta)(71h^2+25\beta^2)\bigg( \mathcal{F}\big(\pi -\frac{\alpha}{2}, \frac{2 \beta}{\beta+h}\big)+ \mathcal{F}\big(\frac{\alpha}{2},\frac{2 \beta}{\beta+h}\big)\bigg),\qquad\qquad\qquad\qquad\qquad\quad\quad
 \\ \nonumber
\Xi_2=
\bigg\{(563h^4+1338h^2 \beta^2+147\beta^4)\bigg( \mathcal{E}\big(\pi -\frac{\alpha}{2}, \frac{2 \beta}{\beta+h}\big)+ \mathcal{E}\big(\frac{\alpha}{2},\frac{2 \beta}{\beta+h}\big)\bigg)\qquad\qquad\qquad\qquad\qquad\\ \nonumber
-8h(h-\beta)(31h^2+33\beta^2)\bigg( \mathcal{F}\big(\pi -\frac{\alpha}{2}, \frac{2 \beta}{\beta+h}\big)+ \mathcal{F}\big(\frac{\alpha}{2},\frac{2 \beta}{\beta+h}\big)\bigg)\bigg\},\qquad\qquad\qquad\qquad\qquad
\\ \nonumber
\Xi_3=\bigg\{4h(1627h^4+6474h^2 \beta^2+2139\beta^4)\bigg( \mathcal{E}\big(\pi -\frac{\alpha}{2}, \frac{2 \beta}{\beta+h}\big)+ \mathcal{E}\big(\frac{\alpha}{2},\frac{2 \beta}{\beta+h}\big)\bigg)\qquad\qquad\qquad\qquad \\ \nonumber
-(h-\beta)(3043h^4+6522h^2\beta^2+675\beta^4)\bigg( \mathcal{F}\big(\pi -\frac{\alpha}{2}, \frac{2 \beta}{\beta+h}\big)+ \mathcal{F}\big(\frac{\alpha}{2},\frac{2 \beta}{\beta+h}\big)\bigg)\bigg\},\qquad\qquad\quad \\ \nonumber
\Xi_4=
\bigg\{(88069h^6+527729h^4\beta^2+34955h^2\beta^4+17787139\beta^6)\bigg( \mathcal{E}\big(\pi -\frac{\alpha}{2}, \frac{2 \beta}{\beta+h}\big)+ \mathcal{E}\big(\frac{\alpha}{2},\frac{2 \beta}{\beta+h}\big)\bigg) \\ \nonumber
-16h(h-\beta)(2689h^4+9662h^2\beta^2+3009\beta^4)\bigg( \mathcal{F}\big(\pi -\frac{\alpha}{2}, \frac{2 \beta}{\beta+h}\big)+ \mathcal{F}\big(\frac{\alpha}{2},\frac{2 \beta}{\beta+h}\big)\bigg)\bigg\}.\qquad\qquad
\\ \nonumber
\end{eqnarray}
Therefore the correction to the lateral Casimir force due to the finite conductivity can be calculated
\begin{eqnarray} \label{conductivity5}
\Delta F_{cc}^{lat} = - \frac{2 \pi}{\lambda} \frac{\partial}{\partial \varphi} \Delta E_{cc}^{cor} =
  \frac{-2 \pi}{\lambda} \frac{A_1 A_2 \sin\varphi}{\sqrt{A_1^2+A_2^2-2 A_1 A_2 \cos\varphi}}\frac{\partial}{\partial \beta} \Delta E_{cc}^{cor}.
\end{eqnarray}
\section{The Finite Temperature Corrections of the Lateral Casimir Force}
The finite temperature Casimir force per unit area for two parallel plates with Dirichlet-Dirichlet or Neumann-Neumann boundary condition is (see \cite{Teo2} and references there in)

\begin{eqnarray} \label{T1}
 \mathfrak{F}_{pp}(h) = -\frac{\pi ^2}{480 h^4}-\frac{\pi ^2 T^4}{90}+\frac{\pi T}{2 h^3} \sum_{k=1}^\infty  \sum_{l=1}^\infty \frac{k^2}{l} \exp(-\frac{\pi k l}{h T}) ,
\end{eqnarray}
or
\begin{eqnarray} \label{T2}
\mathfrak{F}_{pp}(h) = -\frac{\xi_R(3) T}{8 \pi  h^3}-\frac{T}{\pi}\sum_{k=1}^\infty  \sum_{l=1}^\infty \bigg(\frac{2 \pi^2 l^2 T^2}{k h}+\frac{\pi l T}{k^2 h^2}+\frac{1}{4 k^3 h^3}\bigg)e^{-4 \pi h T k l}
\end{eqnarray}

and $h$ is separation distance between the plates. Eq. (\ref{T1}) displays that in the low temperature limit thermal correction is dominated by the term corresponding to $T=0$ and therefore the first term of the thermal corrections in the low temperature region is
\begin{eqnarray} \label{correction}
\triangle\mathfrak{F}_{pp}(h)\sim -\frac{\pi ^2 T^4}{90}.
\end{eqnarray}
Considering this thermal correction and by using PFA, thermal correction of the normal Casimir force reads
\begin{eqnarray} \label{correction of normal f}
\triangle F_{cc}\sim -\frac{L b \pi ^3 T^4}{45}+ \ldots ,
\end{eqnarray}
which is independent of $h$. Corresponding to this force thermal correction of the Casimir energy of two smooth cylinders is
\begin{eqnarray} \label{correction of e}
\triangle E_{cc}\sim \frac{L b \pi ^3 T^4}{45} h+ \ldots .
\end{eqnarray}
Taking sinusoidally corrugation into account, one obtains the following expression for thermal correction of the Casimir energy of the configuration of Fig. (\ref{cylinders}) as
\begin{eqnarray} \label{correction of corrugated}
\triangle E_{cc}^{cor}\sim -\frac{L b \pi ^3 T^4}{45} h+ \ldots .
\end{eqnarray}
There is no phase-dependent in the obtained result. Therefore thermal correction in the low temperature region dose not take part in the lateral Casimir force.
\\
\par

Considering Eq. (\ref{T2}), in the high temperature limit the exponential term goes to zero quickly. Therefore the classical term is the leading term of the thermal correction
\begin{eqnarray} \label{h correction}
\triangle\mathfrak{F}_{pp}(h)\sim -\frac{\xi_R(3) T}{8 \pi  h^3} + \ldots .
\end{eqnarray}
Imposing the proximity force approximation to Eq. (\ref{h correction}) gives
\begin{eqnarray} \label{h correction of normal f}
\triangle F_{cc}\sim -\frac{2 \xi_R(3) T L b}{8 \pi} \int_0^{\pi} \big(\sqrt{b^2+\delta^2-2 b \delta cos\theta}-a \big)^{-3} d\theta + \ldots ,
\end{eqnarray}
where for the case of two interior eccentric cylinders $\delta=b-a-h$.
Notice that since we are in high temperature region, this term may be the main term of the normal Casimir force. Performing the integral we have
\begin{eqnarray} \label{h correction of normal f2}
\triangle F_{cc}\sim -\frac{2 \xi_R(3) T L}{3 \pi^2}\sqrt{\frac{ab}{2(b-a)}}\frac{1}{h^{5/2}} + \ldots ,
\end{eqnarray}
and the thermal correction of the Casimir energy is as follow
\begin{eqnarray} \label{h correction of e}
\triangle E_{cc}\sim -\frac{ \xi_R(3) T L}{16 \pi}\sqrt{\frac{ab}{2(b-a)}} \frac{1}{(h-\beta)\sqrt{h+\beta}}
\bigg\{ \mathcal{E}\big(\pi -\frac{\alpha}{2}, \frac{2 \beta}{\beta+h}\big)+\mathcal{E}\big(\frac{\alpha}{2}, \frac{2 \beta}{\beta+h}\big)\bigg\}+ \ldots .
\end{eqnarray}
Therefore the thermal correction of the lateral Casimir force  between two sinusoidally corrugated eccentric cylinders at high temperature is
\begin{eqnarray} \label{h correction of Lateral}
\triangle F_{cc}^{lat, high T}\sim \frac{ \xi_R(3)\; T L \sqrt{ab}}{16 \sqrt{2(b-a)} \Lambda} \frac{A_1 A_2 sin\varphi}{\sqrt{A_1^2+A_2^2-2 A_1 A_2 cos\varphi}} \Omega(h,\alpha,\beta)+\,
  \ldots ,
\end{eqnarray}
where $\Omega(h,\alpha,\beta)$ is
\begin{eqnarray} \label{omega}
\frac{1}{(h-\beta)^2 \beta (h+\beta)^{3/2}}
\bigg\{(h^2+3 \beta^2)\bigg( \mathcal{E}\big(\pi -\frac{\alpha}{2}, \frac{2 \beta}{\beta+h}\big)+\mathcal{E}\big(\frac{\alpha}{2}, \frac{2 \beta}{\beta+h}\big) \bigg)\\ \nonumber
- h( h- \beta) \bigg( \mathcal{F}\big(\pi -\frac{\alpha}{2}, \frac{2 \beta}{\beta+h}\big)+\mathcal{F}\big(\frac{\alpha}{2}, \frac{2 \beta}{\beta+h}\big) \bigg) \bigg\}.
\end{eqnarray}

\section{Conclusion}
Considering the importance of the Casimir interaction in the nanotechnology, we have studied the lateral Casimir force between two sinusoidally corrugated eccentric cylinders. The mentioned geometry may play an important role as a pair of noncontact gears in micro-electromechanical systems. Applying scattering matrix method results in the exact solution but there exist some difficulties. Therefore we have used the proximity force approximation to achieve our purpose and emitting the difficulties appear in the scattering matrix approach. It is worth mentioning that, one can not use PFA approximation at separations comparable with a period of corrugation. We have assumed the cylinders to be slightly eccentric with similar radiuses and separations much smaller than corrugations' wave length for the validity of PFA. for such short distances the effect of finite conductivity would be non negligible. In addition to the effect of finite conductivity, we have obtained finite-temperature corrections on the lateral Casimir force to reduce the inaccuracy of the result obtained by PFA. Applying this approximation dose not lead in a correction in the low temperature limit but, it concludes in a correction in the hight temperature region. It is worth mentioning that one can obtain the zero-temperature lateral Casimir force between two exterior sinusoidally corrugated cylinders just by replacing $(b-a)$ in Eq.(\ref{F-lateral}) by $(b+a)$.

\section{Acknowledgments}
We thank  A. Moradian for helpful discussions and correspondence. We thank also Kh. Ghamari to design the figure. We would like also to thanks H. Gies,  B. D$\ddot{o}$brich, and  M. Kr$\ddot{u}$ger, for bringing our attentions to the papers \cite{0601094}, \cite{0810.3480}, \cite{1303.2499} respectively.

\nocite{gs}
\nocite{williams}
\nocite{royden}

\bibliographystyle{plain}
\bibliography{martingales}

\end{document}